\RequirePackage{ifpdf}
\ifpdf 
\documentclass[pdftex]{sigma}
\else
\documentclass{sigma}
\fi

\begin{document}
\allowdisplaybreaks

\renewcommand{\PaperNumber}{026}

\newcommand{\njl}{\mathrm{NJL}}
\newcommand{\interaction}{\mathrm{int}}
\newcommand{\crit}{\mathrm{crit}}
\newcommand{\A}{\mathrm{A}}
\newcommand{\h}{\mathrm{H}}
\newcommand{\e}{\mathrm{E}}
\newcommand{\st}{\mathrm{st}}
\newcommand{\pt}{\mathrm{pt}}
\newcommand{\eff}{\mathrm{eff}}
\newcommand{\ud}{\mathrm{d}}
\newcommand{\p}{\mathrm{p}}

\FirstPageHeading

\ShortArticleName{Bosonization of Ef\/fective Multi-Quark
Interactions with $U_\A (1)$ Breaking}

\ArticleName{Functional Integral Approaches to  the Bosonization \\
of Ef\/fective Multi-Quark Interactions\\ with $\boldsymbol{U_\A
(1)}$ Breaking}

\Author{Brigitte HILLER~$^\dag$, Alexander A. OSIPOV~$^\ddag$,
V\'eronique BERNARD~$^\S$ and Alex H. BLIN~$^\dag$}
\AuthorNameForHeading{B. Hiller, A.A. Osipov, V. Bernard and A.H.
Blin}

\Address{$^\dag$~Centro de F\'{\i}sica Te\'{o}rica,
Departamento de
            F\'{\i}sica da Universidade de Coimbra,\\
            $\phantom{^\dag}$~3004-516 Coimbra, Portugal}

\EmailD{\href{mailto:brigitte@teor.fis.uc.pt}{brigitte@teor.fis.uc.pt},
 \href{mailto:alex@teor.fis.uc.pt}{alex@teor.fis.uc.pt}}

\Address{$^\ddag$~Laboratory of Nuclear
            Problems, Joint Institute for Nuclear Research,\\
            $\phantom{^\ddag}$~141980 Dubna, Moscow Region, Russia}
\EmailD{\href{mailto:osipov@nusun.jinr.ru}{osipov@nusun.jinr.ru}}

\Address{$^\S$~Laboratoire de Physique Th\'eorique 3-5,
Universit\'e Louis Pasteur,
             rue de l'Universit\'e,\\
            $\phantom{^\S}$~F-67084 Strasbourg, France}
\EmailD{\href{mailto:bernard@lpt6.u-strasbg.fr}{bernard@lpt6.u-strasbg.fr}}

\ArticleDates{Received October 27, 2005, in f\/inal form February
13, 2006; Published online February 23, 2006}

\Abstract{Low energy hadron phenomenology involving the (u,d,s)
quarks is often approached through ef\/fective multi-quark
Lagrangians with the symmetries of QCD. A very successful approach
consists in taking the four-quark Nambu--Jona-Lasinio Lagrangian
with the chiral $U_L(3)\times U_R(3)$ symmetry in the massless
limit, combined with the $U_\A(1)$ breaking six-quark f\/lavour
determinant interaction of 't Hooft. We review the present status
and some very recent developments related to the functional
integration over the cubic term in auxi\-liary mesonic variables
that one introduces to bosonize the system. Various approaches for
handling this functional, which cannot be integrated exactly, are
discussed: the statio\-nary phase approximation, the perturbative
expansion, the loop expansion, their interrelation and importance
for the evaluation of the ef\/fective action. The intricate group
structure rules out the method of Airy's integral. The problem of
the instability of the vacuum is stated and a solution given by
including eight-quark interactions.}


\Classification{81T10}

\section{Introduction}

Non-perturbative methods to deal with the low energy regime of QCD
involve lattice calcula\-tions, QCD sum rules \cite{Shifman:1979},
ef\/fective f\/ield theories, which are renormalizable in the
modern sense and gave birth to Chiral Perturbation Theory
\cite{Weinberg:1979,Gasser:1983}, large $N_c$ (number of colours)
expansions~\cite{Hooft:1974,Witten:1979b,Manohar:1998,Pich:2002},
and ef\/fective models with solid symmetry content
\cite{Gasiorowicz:1969}. The object of our studies is placed in
the latter category, consisting in a model Lagrangian which
combines the four-quark interactions of Nambu and Jona-Lasinio
(NJL) \cite{Nambu:1961}, extended to the $U_L(3)\times U_R(3)$
chiral symmetry of massless QCD
\cite{Eguchi:1976,Volkov:1982,Ebert:1986}, and the six-quark
interaction of 't Hooft \cite{Hooft:1978}, which breaks the
unwanted $U_\A(1)$ symmetry of QCD. We call it the NJLH
Lagrangian. The $2N_f$ ($N_f$ denoting the number of f\/lavours)
multi-quark interactions of 't Hooft were derived after
eliminating the gluonic degrees of freedom of QCD in the
semiclassical instanton approximation, assuming dominance of the
zero modes in the scattering process of quarks and anti-quarks on
the same instanton. Recent reviews providing further evidence on
multi-quark vertices can be found in~\cite{Diakonov:2003}.

In the last decade some new developments brought further relevance
to the NJLH model, as it may help to shed some light on the
possible existence of an hierarchy of multi-quark interactions,
with prevalence of lower ones. There is some conf\/lict on two
fronts: (i) Going beyond the zero mode approximation, an
inf\/inite number of multi-quark interactions, starting from the
four-quark ones, has been found in the instanton-gas model, all of
them being of the same importance \cite{Simonov:1997}. (ii) On the
other hand, accurate lattice measurements for the realistic QCD
vacuum show a hierarchy between the gluon f\/ield correlators with
a dominance of the lowest one \cite{Bali:2001}. A similar
hierarchy of the multi-quark interactions can be triggered after
averaging over gluon f\/ields. If this is true, there is an
apparent contradiction with the instanton-gas model (see also
\cite{Witten:1979}). This point has been stressed in
\cite{Simonov:2002}.

The hierarchy problem of multi-quark interactions can be addressed
on pure phenomeno\-logical grounds. For that we suggest to
simplify the task considering only the four- and six-quark
interactions of the NJLH model. Once one knows the Lagrangian the
obvious question arises: does the system possess a stable vacuum
state and does this state correspond to our phenomenological
expectations? If hierarchy takes place this question is pertinent
for the leading four-quark interaction, because in this case the
ef\/fective quark Lagrangian can be studied step by step in the
hierarchy with the assumption that four-quark vertices are the
most important ones. In the opposite case we must study the system
as a whole to answer the question. For the best known and simplest
example, to which we dedicate most of our attention here, the
solution may be found analytically. As a result one can obtain
def\/inite answers to the above questions with a convincing
indication in favour of a hierarchy for the considered example.

Pioneering work on the original NJLH Lagrangian has been done in
\cite{Bernard:1988}, using Bethe--Salpeter techniques to evaluate
quark -- anti-quark scattering in one-loop approximation and
obtain the meson masses from the poles of the scattering matrix,
and in \cite{Reinhardt:1988},  where the quark Lagrangian has been
bosonized using functional integral methods and the mesonic
spectrum calculated at the leading order of the stationary phase
approach (SPA) for one critical point.  A good description of the
pseudoscalar nonet, especially the $\eta$ and $\eta'$ masses and
mixing has been achieved. Within the respective approximations,
the ef\/fective potentials for both methods
coincide~\cite{Osipov:2004}. The model has subsequently been
widely and successfully explored at the mean f\/ield level, see
e.g.~\cite{Klimt:1990,Bernard:1993,Hatsuda:1994}.

In the present work we present a detailed picture and
mathematically careful study of the functional integration over
the auxiliary bosonic f\/ields of cubic order, inherited from the
't Hooft six-quark interaction. They cannot be integrated out
exactly. We came across one startling fact and several interesting
properties, which we hope help in the understanding of the
approximations used and in the identif\/ication of the
shortcomings and potentialities of multi-quark models. Before
turning to the calculations, let us summarize what we have learned
from them~\cite{Osipov:2005a,Osipov:2005}:

(1) We show that the stationary phase equations have several roots
(critical points), of which one is regular and the others
singular. A rigorous SPA treatment, taking into account all
critical points, leads to an unstable vacuum for the theory. This
is the startling point, since this fatal f\/law of the model is in
crass contradiction with its phenomenological success.

(2) The results in \cite{Reinhardt:1988} are obtained if one
considers only the regular critical point of the SPA, which of
course is accompanied with a dif\/ferent asymptotic behaviour.

(3) This result can be understood in the perturbative sense, where
the cubic interaction is taken as a perturbation around the stable
NJL vacuum. The situation is in many respects analogous to the
problem of a harmonic oscillator perturbed by an $x^3$ term. This
system has no ground state, but perturbation theory around a local
minimum does not know this.

(4) By performing a resummation of the perturbative series in
powers of the 't Hooft interaction one gets a loop expansion
\cite{Coleman:1973}, of which the leading term contains all tree
graphs present in all powers of the perturbative series. This
corresponds exactly to the result of \cite{Reinhardt:1988}.

(5) The loop expansion can be classif\/ied in terms which
contribute to the measure (all odd numbers of loops) and to the
phase, relevant for the ef\/fective action (all even number of
loops).

(6) The global instability of the vacuum can be removed within the
multi-quark interaction picture: the addition of a chiral
invariant OZI-violating \cite{Okubo:1963} eight-quark interaction
to the Lagrangian renders it globally stable, for certain values
of the coupling strengths of the several interactions.

(7) A comparison between the ef\/fective potentials for the NJLH
and for the eight-quark enlarged Lagrangian reveals that the
global stabilization happens around the local minimum which
appears if only the regular critical point is considerd in the
NJLH Lagrangian. This is probably at the heart of its success.

Points (6) and (7) will not be discussed in this contribution. In
fact, at the time the conference took place, they had not been
obtained yet. We added them to the list, as they represent a kind
of ``happy-end'' for the model. We refer to \cite{Osipov:2005} for
a detailed presentation of this issue.

The paper is organized as follows. In Section 2 we write out the
NJLH Lagrangian and use the functional integral representation to
express it in terms of bosonic f\/ields. In Section 3 we show that
the chiral symmetry group imposes constraints which are only
compatible either with the perturbative approach, or the expansion
in a parameter that multiplies the total Lagrangian density (the
loop expansion). Otherwise, as the consistent stationary phase
treatment shows, the model is unstable. In Section 4.1 we discuss
the perturbative treatment; the integration over auxiliary
variables leads to a special problem with $\delta
(0)\,$-singularities, to which we give a physical meaning through
Feynman diagrams in 4.2. The loop expansion is considered in
Section 5. We obtain in closed form the two-loop contributions to
the functional integral ${\cal Z}$ and give arguments to justify
this result. Since we have forestalled the conclusions in the
Introduction, we shall omit concluding remarks.

\section{The Lagrangian and bosonization}

On lines suggested by multicolour chromodynamics it can be argued
\cite{Witten:1979a} that the $U_\A (1)$ anomaly vanishes in the
large $N_c$ limit, so that mesons come degenerate in mass nonets.
Hence the leading order (in $N_c$ counting) mesonic Lagrangian and
the corresponding underlying quark Lagrangian must inherit the
$U_L(3)\times U_R(3)$ chiral symmetry of massless QCD. In
accordance with these expectations the $U_L(3)\times U_R(3)$
symmetric NJL interactions,
\begin{gather}
\label{L4q}
  {\cal L}_\njl =\frac{G}{2}\left[(\bar{q}\lambda_aq)^2+
                 (\bar{q}i\gamma_5\lambda_aq)^2\right],
\end{gather}
can be used to specify the corresponding local part of the
ef\/fective quark Lagrangian in channels with quantum numbers
$J^P=0^+, 0^-$. The Gell-Mann matrices and the singlet
$\lambda_0=\sqrt{\frac{2}{3}}\ \mbox{\Large 1}$ act in f\/lavour
space and are denoted by $\lambda_a$, $a=0,1,\ldots ,8,$ obeying
the basic property $\mbox{tr}\, \lambda_a \lambda_b
=2\delta_{ab}$.

The 't Hooft determinantal interactions are described by the
Lagrangian \cite{Hooft:1978}
\begin{gather}
\label{Ldet}
  {\cal L}_{\h}=\kappa (\det \bar{q}P_Lq
                         +\det \bar{q}P_Rq),
\end{gather}
where the matrices $P_{L,R}=(1\mp\gamma_5)/2$ are projectors and
the determinant is over f\/lavour indices.

The coupling constant $\kappa$ is a dimensional parameter
($[\kappa ]=\mbox{GeV}^{-5}$) with the large $N_c$ asymptotics
$\kappa\sim 1/N_c^{N_f}$. The coupling $G$, $[G]=\mbox{GeV}^{-2}$,
counts as $G\sim 1/N_c$ and, therefore, the Lagrangian (\ref{L4q})
dominates over ${\cal L}_\h$ at large $N_c$. It dif\/fers from the
counting $G\sim 1/N_c^2$, which one obtains in the instanton-gas
vacuum \cite{Simonov:1997}.

It is assumed here for simplicity that interactions between quarks
can be taken in the long wavelength limit where they are
ef\/fectively local. The 't Hooft-type ansatz (\ref{Ldet}) is a
frequently used approximation.  Even in this essentially
simplif\/ied form the determinantal interaction has all basic
ingredients to describe the dynamical symmetry breaking of the
hadronic vacuum and explicitly breaks the axial $U_\A (1)$
symmetry \cite{Diakonov:1996,Dorokhov:1992}. The ef\/fective
mesonic Lagrangian in leading order SPA, corresponding to the
non-local determinantal interaction, has been found in
\cite{Diakonov:1985,Simonov:1995}.

Anticipating our result, we would like to note that if the
hierarchy of multi-quark interactions really occurs in nature, the
perturbative treatment seems adequate. The NJL interaction alone
has a stable vacuum state corresponding to spontaneously broken
chiral symmetry. But, as we shall show, the ef\/fective quark
theory based on the Lagrangian\footnote{The other multi-quark
terms have been neglected here.}
\begin{gather}
\label{totlag}
  {\cal L}_{\rm NJLH}=\bar{q}(i\gamma^\mu\partial_\mu -\hat{m})q
          +{\cal L}_\njl + {\cal L}_\h
\end{gather}
has a fatal f\/law: if ${\cal L}_\h$ is comparable with ${\cal
L}_\njl$, it has no stable ground state.  In
equation~(\ref{totlag}) the current quark mass, $\hat{m}$, 
stands for the diagonal matrix  $\mbox{diag}\, (\hat{m}_u, \hat{m}_d, \hat{m}_s)$, 
which explicitly breaks the global chiral $SU_L(3)\times SU_R(3)$ symmetry of the Lagrangian.

To bosonize the theory one introduces auxiliary bosonic variables
to render fermionic vertices bilinear in the quark f\/ields. This
procedure requires twice more bosonic degrees of freedom than
necessary~\cite{Reinhardt:1988}. Redundant variables must be
integrated out and this integration is problematic as soon as one
goes beyond the lowest order stationary phase approximation
\cite{Osipov:2002,Osipov:2004}: the lowest order result is simply
the value of the integrand taken at one def\/inite stationary
point \cite{Diakonov:1996}. We shall show how to extract
systematically the higher order corrections which contribute to
the ef\/fective mesonic Lagrangian, and what to do with
inf\/inities contained in these corrections.

The many-fermion vertices of Lagrangian (\ref{totlag}) can be
linearized by introducing the functional unity
\cite{Reinhardt:1988}
\begin{gather}
  1 = \int \prod_a {\cal D}s_a{\cal D}p_a\ \delta
                    (s_a-\bar{q}\lambda_aq)
                    \delta (p_a-\bar{q}i\gamma_5\lambda_aq)
                    \nonumber \\
\phantom{1}{} = \int \prod_a {\cal D}s_a {\cal D}p_a
                    {\cal D}\sigma_a {\cal D}\phi_a
                    \exp \left\{i\int\!\ud^4x
                    \left[\sigma_a(s_a-\bar{q}\lambda_aq) +
                    \phi_a(p_a-\bar{q}i\gamma_5\lambda_aq)\right]
                    \right\}\label{1}
\end{gather}
in the vacuum-to-vacuum amplitude
\begin{gather}
\label{genf1}
   Z=\int {\cal D}q{\cal D}\bar{q}\ \exp\left(i\int
     \ud^4x{\cal L}_{\rm NJLH}\right).
\end{gather}
We consider the theory of quark f\/ields in four-dimensional
Minkowski space. It is assumed that the quark f\/ields have colour
$(N_c=3)$ and f\/lavour $(N_f=3)$. The auxiliary bosonic
f\/ields,~$\sigma_a$, and, $\phi_a$, $(a=0,1,\ldots, 8)$ become
the composite scalar and pseudoscalar mesons and the auxiliary
f\/ields,~$s_a$, and, $p_a$, must be integrated out.

By means of the simple trick (\ref{1}), it is easy to write down
the amplitude~(\ref{genf1}) as
\begin{gather}
\label{genf2}
   Z=\int {\cal D}q{\cal D}\bar{q}
     \prod^8_{a=0}{\cal D}s_a
     \prod^8_{a=0}{\cal D}p_a
     \prod^8_{a=0}{\cal D}\sigma_a
     \prod^8_{a=0}{\cal D}\phi_a
     \exp\left(i\int \ud^4x{\cal L}'\right)
\end{gather}
with
\begin{gather*}
      {\cal L}'= \bar{q}(i\gamma^\mu\partial_\mu -\sigma
               - i\gamma_5\phi )q + s_a (\sigma_a -\hat{m}_a)
               + p_a\phi_a + {\cal L}_\njl' + {\cal L}_\h' ,
\\
      {\cal L}_\njl' = \frac{G}{2}\left[(s_a)^2+(p_a)^2\right] ,
\\
      {\cal L}_\h' = \frac{\kappa}{64}\left[ \det(s+ip)
                     +\det (s-ip)\right]
                   = \frac{\kappa}{32}A_{abc}s_a
                     \left(s_bs_c-3p_bp_c \right).
\end{gather*}
We assume here that $\sigma =\sigma_a\lambda_a$, and so on for all
auxiliary f\/ields $\sigma ,\phi ,s,p$.

The totally symmetric constants $A_{abc}$ are related to the
f\/lavour determinant, and equal to
\begin{gather*}
   A_{abc}
 =\frac{1}{3!}\epsilon_{ijk}\epsilon_{mnl}(\lambda_a)_{im}
             (\lambda_b)_{jn}(\lambda_c)_{kl} \nonumber \\
\phantom{A_{abc}}{}=\frac{2}{3}d_{abc} +
      \sqrt{\frac{2}{3}} \big(
      3\delta_{a0}\delta_{b0}\delta_{c0}
      -\delta_{a0}\delta_{bc}
      -\delta_{b0}\delta_{ac}
      -\delta_{c0}\delta_{ab}\big).
\end{gather*}
We use the standard def\/initions for antisymmetric $f_{abc}$ and
symmetric $d_{abc}$ structure constants of $U(3)$ f\/lavour
symmetry. One can f\/ind, for instance, the following useful
relations
\begin{gather*}
 f_{eac}A_{bfc}+f_{ebc}A_{fac}+f_{efc}A_{abc}=0, \nonumber \\
 d_{eac}A_{bfc}+d_{ebc}A_{fac}+d_{efc}A_{abc}
     = \sqrt{6}\delta_{e0}A_{abf}, \nonumber \\
\sum_{b=0}^8 A_{abb} = -2 \sqrt{\frac{2}{3}}\, \delta_{a0},
     \qquad \sum_{c,e=0}^{8} A_{ace}A_{bce} = \frac{8}{9}\, \delta_{ab}.
\end{gather*}

At this stage it is easy to rewrite equation~(\ref{genf2}), by
changing the order of integrations, in a~form appropriate to
accomplish the bosonization, i.e., to calculate the integrals over
quark f\/ields and integrate out from $Z$ the unphysical part
associated with the auxiliary bosonic variables~$(s_a, p_a)$
\begin{gather}
   Z =\int \prod_a{\cal D}\sigma_a{\cal D}\phi_a
                    {\cal D}q{\cal D}\bar{q}\,
                    \exp\left(i\int\ud^4x
                    {\cal L}_q(\bar{q},q,\sigma ,\phi )\right)
                    \nonumber \\
 \phantom{Z =}  {}\times \int \prod_a{\cal D}s_a{\cal D}p_a\,
    \exp\left(i\int\ud^4x{\cal L}_r(\sigma ,\phi ,s,p)\right),\label{genf3}
\end{gather}
where
\begin{gather*}
  {\cal L}_q  =
  \bar{q}(i\gamma^\mu\partial_\mu -\sigma - i\gamma_5\phi )q, \\
  {\cal L}_r  =  s_a(\sigma_a - \hat{m}_a) + p_a\phi_a
  + {\cal L}'_\njl + {\cal L}'_\h .
\end{gather*}
The Fermi f\/ields enter the action bilinearly, thus one can
always integrate over them, since one deals with a Gaussian
integral. One should also shift the scalar f\/ields
$\sigma_a(x)\rightarrow\sigma_a(x)+m_a$ by demanding that the
vacuum expectation values of the shifted f\/ields vanish $\big <
0|\sigma_a(x)|0\big >=0$. In other words, all tadpole graphs in
the end should sum to zero, giving us the gap equation to f\/ix
the constituent quark masses $m_a$ corresponding to the physical
vacuum state.

The functional integrals over $s_a$ and $p_a$
\begin{gather}
\label{intJ}
     {\cal Z}[\sigma ,\phi ;\Delta ]\equiv
     {\cal N}\int^{+\infty}_{-\infty}\prod_a{\cal D}s_a{\cal D}p_a\,
     \exp\left(i\int\ud^4x{\cal L}_r(\sigma +m,\phi ,s,p)\right)
\end{gather}
are the main subject of our study. We put here
$\Delta_a=m_a-\hat{m}_a$, and ${\cal N}$ is chosen so that ${\cal
Z}[0,0;\Delta ]=1$.

Let us join the auxiliary bosonic variables in one $18$-component
object $R_A=(R_a,R_{\dot{a}})$ where we identify $R_a\equiv s_a$
and $R_{\dot{a}}\equiv p_a$; $a$, $\dot{a}$ run from $0$ to $8$
independently. It is clear then, that $R_A^2=s_a^2+p_a^2$.
Analogously, we will use $\Pi_A=(\sigma_a,\phi_a)$ for external
f\/ields and $\Delta_A=(\Delta_a,0)$.

Next, consider the sum $\Phi_{ABC}R_AR_BR_C$. If we require
\begin{gather*}
   \Phi_{abc}=\frac{3}{16}A_{abc} , \qquad
   \Phi_{a\dot{b}\dot{c}}=-\frac{3}{16}A_{abc} , \qquad
   \Phi_{ab\dot{c}}=\Phi_{\dot{a}\dot{b}\dot{c}}=0,
\end{gather*}
we f\/ind after some algebra
\begin{gather*}
   \frac{\kappa}{3!}\ \Phi_{ABC}R_AR_BR_C = {\cal L}_\h'
\end{gather*}
with the following important property to be fulf\/illed
\begin{gather}
\label{contr}
      \Phi_{ABC}\delta_{BC}=0.
\end{gather}

Now it is easy to see that the functional integral (\ref{intJ})
can be written in a compact way
\begin{gather}
\label{intJi}
     {\cal Z}[\Pi ,\Delta ]\equiv
     {\cal N}\int^{+\infty}_{-\infty}\prod_A{\cal D}R_A\,
     \exp\left(i\int\ud^4x{\cal L}_r(\Pi ,\Delta ;R)\right),
\end{gather}
where
\begin{gather}
\label{Lr}
   {\cal L}_r = R_A (\Pi_A+\Delta_A) + \frac{G}{2} R_A^2
   + \frac{\kappa}{3!}\ \Phi_{ABC}R_AR_BR_C.
\end{gather}
We have arrived at a functional integral with a cubic polynomial
in the exponent.

\section[The integration over the auxiliary variables $R_A$]{The integration over the auxiliary
variables $\boldsymbol{R_A}$}

Given the cubic structure in the functional integral, one might be
tempted to solve it using Airy's integral methods. Invoking the
existence of a large expansion parameter, such as $N_c$, one uses
the well known asymptotics of the Airy's function on the real
axis. This would require \cite{Osipov:2005a}
\begin{gather}
\label{sdLr}
   \frac{\partial^2 {\cal L}_r}{\partial R_A \partial R_B}
   \equiv {\cal L}_{AB}'' =
   G\delta_{AB} + \kappa \Phi_{ABC}R_C = 0
\end{gather}
which cannot be fulf\/illed.

On the contrary, the system of equations based on the f\/irst
order derivatives
\begin{gather}
\label{fdLr}
   \frac{\partial {\cal L}_r}{\partial R_A} =
   GR_A + \Delta_A + \Pi_A + \frac{\kappa}{2}\ \Phi_{ABC}R_BR_C = 0
\end{gather}
is self-consistent and can be solved \cite{Osipov:2004}.
Therefore, one can obtain the semi-classical asymptotics through
the stationary phase method.

\subsection[Solving equation (13)]{Solving equation (\ref{fdLr})}

We need to recall shortly the solutions of equation~(\ref{fdLr}).
Up to some order in the external mesonic f\/ields, $\Pi_A$, we may
write them as a polynomial $R_A={\cal R}_A^{(i)}$
\begin{gather*}
   {\cal R}_A^{(i)} = H_A^{(i)}
                    + H_{AB}^{(i)}\Pi_B
                    + H_{ABC}^{(i)}\Pi_B\Pi_C
                    + H_{ABCD}^{(i)}\Pi_B\Pi_C\Pi_D
                    + \cdots ,
\end{gather*}
where $i=1,2,\ldots$ denote dif\/ferent possible solutions. The
coef\/f\/icients $H_{A\ldots}^{(i)}$ depend on $\Delta_a$ and on
the coupling constants $G$, $\kappa$, and the higher index
coef\/f\/icients $H_{A\ldots}^{(i)}$ are recurrently expressed in
terms of the lower ones. For instance, we have
\begin{gather*}
   \left(H_{AB}^{(i)}\right)^{-1} = -\left(G\delta_{AB}
                                    + \kappa\Phi_{ABC}H_C^{(i)}\right),
\\
   H_{ABC}^{(i)} = \frac{\kappa}{2}\, \Phi_{DEF} H_{DA}^{(i)} H_{EB}^{(i)}
                   H_{FC}^{(i)} ,
\end{gather*}
and so on.

Putting this expansion in equation~(\ref{fdLr}) one obtains a
series of self-consistent equations to determine
$H_{A\ldots}^{(i)}$. The f\/irst one is
\begin{gather*}
   GH_A^{(i)} +\Delta_A+\frac{\kappa}{2}\ \Phi_{ABC}
   H_B^{(i)}H_C^{(i)} = 0.
\end{gather*}

One can always f\/ind the trivial solution $H_A=0$, corresponding
to the unbroken vacuum $\Delta_A=0$. There are also non-trivial
ones for the scalar component, i.e.,
\begin{gather}
\label{HA}
   H_A^{(i)} = \big(h_a^{(i)},0\big).
\end{gather}

The number of possible solutions, $i$, depends on the symmetry
group. The coef\/f\/icients $h_a^{(i)}$ are determined by the
couplings $G$, $\kappa$ and the mean f\/ield value $\Delta_a$. In
accordance with the pattern of explicit symmetry breaking the mean
f\/ield can have only three non-zero components at most with
indices $a=0,3,8$. If two of 
the three indices in $A_{abc}$ are from the set $\{0,3,8\}$, then the third one also belongs to this set. Thus
$\Delta_a$ is the only object which determines the vector
structure of the solution $h_a^{(i)}$, and therefore
$h_a^{(i)}\neq 0$ if $a=0,3,8$. It means that in general we have a
system of only three equations to determine
$h^{(i)}=h_a^{(i)}\lambda_a=\mbox{diag}\,(h_u^{(i)},h_d^{(i)},h_s^{(i)})$
\begin{gather}
Gh_u+\Delta_u+\displaystyle\frac{\kappa}{16}h_dh_s=0,\nonumber \\
 Gh_d+\Delta_d+\displaystyle\frac{\kappa}{16}h_sh_u=0,\nonumber \\
Gh_s+\Delta_s+\displaystyle\frac{\kappa}{16}h_uh_d=0 .
 \label{saddle-1}
\end{gather}

This system is equivalent to a f\/ifth order equation for a
one-type variable which can be solved numerically. For two
particular cases, when $\hat{m}_u=\hat{m}_d=\hat{m}_s$ and
$\hat{m}_u=\hat{m}_d\neq\hat{m}_s$, equations~(\ref{saddle-1}) can
be solved analytically \cite{Osipov:2004}. The simplest example:
$\hat{m}_u=\hat{m}_d=\hat{m}_s$ (or, equivalently,
$h_u^{(i)}=h_d^{(i)}=h_s^{(i)}$) corresponds to $SU(3)$ f\/lavour
symmetry. In this case equation~(\ref{saddle-1}) has two solutions
\begin{gather}
\label{hsu3}
   h_u^{(1)} = -\frac{8G}{\kappa} \left( 1-
   \sqrt{1-\frac{\kappa\Delta_u}{4G^2}} \right)\ , \qquad
   h_u^{(2)} = -\frac{8G}{\kappa} \left( 1+
   \sqrt{1-\frac{\kappa\Delta_u}{4G^2}} \right).
\end{gather}
If $4G^2>\kappa\Delta$, they are real and will contribute to the
stationary phase trajectory.

\subsection{The lowest order semiclassical
                                asymptotics and instability}

Since the system of equations (\ref{fdLr}) can be solved, we may
replace variables $R_A\rightarrow \bar{R}_A=R_A-{\cal R}_A^{(i)}$
in the functional integral (\ref{intJi}) to obtain the
semi-classical asymptotics
\begin{gather}
     {\cal Z}[\Pi ,\Delta ] \sim
     {\cal N} \sum_{i=1}^n
     \exp\left( i\int\ud^4x {\cal L}_\st^{(i)}\right)
     \int\limits^{+\infty}_{-\infty}\prod_A{\cal D}\bar{R}_A\,
     \exp\left(\frac{i}{2}\int\ud^4x{\cal L}_{AB}''({\cal R}^{(i)})
     \bar{R}_A \bar{R}_B \right) \nonumber \\
  \phantom{{\cal Z}[\Pi ,\Delta ] \sim}{}  \times \sum_{k=0}^\infty \frac{1}{k!}
     \left(i\frac{\kappa}{3!}\, \Phi_{ABC}\int\ud^4x
     \bar{R}_A\bar{R}_B\bar{R}_C\right)^k \qquad  (\hbar\to 0),\label{intJisp}
\end{gather}
where $n$ is the number of solutions, ${\cal R}^{(i)}_A$, of
equation~(\ref{fdLr}), ${\cal L}_{AB}''$ has been def\/ined in
equation~(\ref{sdLr}), and
\begin{gather*}
   {\cal L}_\st^{(i)} =
                {\cal R}^{(i)}_A (\Pi_A+\Delta_A)
                + \frac{G}{2} \left({\cal R}^{(i)}_A\right)^2
                + \frac{\kappa}{3!}\ \Phi_{ABC}
                {\cal R}^{(i)}_A{\cal R}^{(i)}_B{\cal R}^{(i)}_C
                \nonumber \\
\phantom{{\cal L}_\st^{(i)}}{}   =\frac{G}{6}\left({\cal
R}_A^{(i)}\right)^2\!
                +\frac{2}{3}\ {\cal R}^{(i)}_A \left(\Pi_A
                + \Delta_A\right)
                = h_a^{(i)}\sigma_a + {\cal O}\big(\Pi^2\big).
\end{gather*}
Here we used equation~(\ref{fdLr}) to eliminate the term
proportional to $\kappa$. Let us stress that ${\cal L}_\st^{(i)}$
depends on $\kappa$ implicitly: $\kappa$ is contained in ${\cal
R}^{(i)}_A$, or more precisely in the coef\/f\/icients
$H^{(i)}_{A...}$ which are functions of $h^{(i)}_a$. This
dependence is singular at $\kappa\to 0$. One can see this, for
instance, from equation~(\ref{hsu3}) where $h_u^{(2)}\to\infty$
for small $\kappa$. This behaviour ref\/lects the fact that we are
far from the perturbative regime, meaning that the interactions
${\cal L}_\njl$ and ${\cal L}_\h$ are equally weighted.

The linear term in the $\sigma$ f\/ield is written explicitly.
This part of the Lagrangian is responsible for the dynamical
symmetry breaking in the multi-quark system and taken together
with the corresponding part from the Gaussian integration over
quark f\/ields in equation~(\ref{genf3}) leads us to the gap
equation.

At leading order, $k=0$, and for the terms linear in the f\/ields
contributing to the phase of ${\cal Z}$ we have the estimate for
equation~(\ref{intJisp})
\begin{gather*}
     {\cal Z}\sim\sum_{i=1}^n A_{(i)}
     \exp\left( i\int\ud^4x {\cal L}_\st^{(i)}\right)
     \sim \exp \left(i\int\ud^4x \frac{1}{n}\sum_{i=1}^n
     h_a^{(i)}\sigma_a + \cdots\right),
\end{gather*}
where $A_{(i)}$ is real and proportional to\footnote{The term
constant in the f\/ields of $A_{(i)}$ is identical for both
critical points.}
\begin{gather*}
   A_{(i)} \sim |\det {\cal L}_{AB}''({\cal R}^{(i)})|^{-1/2}.
\end{gather*}

Therefore, if one considers the case with
$\hat{m}_u=\hat{m}_d=\hat{m}_s$, ${\cal Z}$ is given by
\begin{gather}
   {\cal Z} \sim
           \exp\left(i\int\ud^4x \frac{1}{2}\sum_{i=1}^2
           h_a^{(i)}\sigma_a + \cdots\right)
            \sim \exp\left(-i\frac{4G}{\kappa}
           \int\ud^4x (\sigma_u + \sigma_d + \sigma_s ) + \cdots\right).\label{gec1}
\end{gather}

Let us recall that the quark loop contribution to the gap equation
is well known (see, for instance, \cite{Osipov:2004NPA}).
Combining this known result with the estimate (\ref{gec1}), one
can obtain the corresponding ef\/fective potential $U(m)$ as a
function of the constituent quark mass $m$
\begin{gather}
\label{effpot}
     U(m)  = \frac{12G}{\kappa}\, m
     -\frac{3N_c}{16\pi^2}\left[m^2\left(\Lambda_q^2
     - m^2\ln\left(1+\frac{\Lambda_q^2}{m^2}\right)\right)
   +
     \Lambda_q^4\ln\left(1+\frac{m^2}{\Lambda_q^2}\right)\right],
\end{gather}
where we consider the case $\hat{m}=0$ for simplicity and
$\Lambda_q$ denotes the cutof\/f of quark loop integrals. This
system has at most a metastable vacuum state (for $G/\kappa >0$);
for $G/\kappa <0$, the ef\/fective potential does not have extrema
in the region $m>0$. We must conclude that the model considered
has a fatal f\/law and can be used only in the framework of the
perturbative approach, which does assume the hierarchy of
multi-quark interactions.

\section[Perturbative expansion of ${\cal Z}$]{Perturbative expansion of $\boldsymbol{\cal Z}$}

We shall restrict ourselves in this section to the perturbative
treatment of the functional integ\-ral~(\ref{intJi}). The loop
expansion will be considered in the next section.

\subsection{The perturbative series}

Let us divide the Lagrangian (\ref{Lr}) in two parts. The free
part, ${\cal L}_0$, is given by
\begin{gather*}
   {\cal L}_0 (R_A) =
   \frac{G}{2}\, R_A^2 + R_A(\Pi_A +\Delta_A).
\end{gather*}
The 't Hooft interaction is considered as a perturbation ${\cal
L}_I$
\begin{gather*}
   {\cal L}_I (R_A) =
   \frac{\kappa}{3!}\, \Phi_{ABC}R_AR_BR_C.
\end{gather*}

Thus the perturbative representation for the functional integral
(\ref{intJi}) can be written as
\begin{gather}
\label{Zpt1}
   {\cal Z}= N'\,\exp\left( i\int\ud^4x {\cal L}_I (\hat{X}_A) \right)
   \int\prod_{A} {\cal D} R_A\ e^{i \int\ud^4x {\cal L}_0 (R_A)},
\end{gather}
where
\begin{gather*}
   \hat{X}_A = -i\,\frac{\delta}{\delta\Pi_A} .
\end{gather*}

Since the boson f\/ields appear quadratically in
equation~(\ref{Zpt1}), they may be integrated out, yielding
\begin{gather}
\label{Zpt2}
   {\cal Z}= N\,\exp\left( i\int\ud^4x {\cal L}_I (\hat{X}_A) \right)
     \exp \left(-i \int\ud^4x
     \frac{\bar{\Pi}_A^2}{2G}\right),
\end{gather}
where $\bar{\Pi}_A=\Pi_A + \Delta_A$. The overall factor $N=(-2\pi
i/G)^9N'$ is unimportant in the following.

We want to calculate the ef\/fective action
$\Gamma_{\eff}=\int\ud^4x {\cal L}_{\eff}$, which by def\/inition
is the phase of ${\cal Z}$
\begin{gather}
\label{Zpt3}
   {\cal Z}=A(\bar\Pi_A )\,\exp \left(i\Gamma_{\eff}
   (\bar\Pi_A) \right),
\end{gather}
and $A(\bar\Pi_A )$ is a real function. Comparing (\ref{Zpt2}) and
(\ref{Zpt3}), one gets
\begin{gather}
\label{Gamma_eff}
   \Gamma_{\eff}=i\ln\frac{A}{N}+\Gamma_0
   -i\ln\left(1+e^{-i\Gamma_0}
   \left( e^{i\int\ud^4x {\cal L}_I}-1\right) e^{i\Gamma_0}\right).
\end{gather}
Here $\Gamma_0$ represents the leading order result for
$\Gamma_\eff$
\begin{gather}
\label{gamma0}
   \Gamma_0 = - \frac{1}{2G}\int\ud^4x \bar\Pi_A^2
\end{gather}
while the second logarithm in equation~(\ref{Gamma_eff}) is a
source of $U_\A (1)$ breaking corrections which arise as a series
in powers of the functional derivatives operator
\begin{gather*}
   \hat\Gamma_I = \int\ud^4x {\cal L}_I(\hat{X}_A)
   = \frac{\kappa}{3!}\,\Phi_{ABC}\int\ud^4x \hat{X}_A \hat{X}_B \hat{X}_C .
\end{gather*}

To make this statement more explicit let us consider the expansion
\begin{gather*}
   \delta = e^{-i\Gamma_0}
   \left( e^{i\hat\Gamma_I}-1 \right) e^{i\Gamma_0} =
   \sum_{m=1}^\infty \frac{i^m}{m!}\left(
   e^{-i\Gamma_0}\hat\Gamma_I e^{i\Gamma_0} \right)^m.
\end{gather*}
Taking into account the symmetry properties of the
coef\/f\/icients $\Phi_{ABC}$ and our previous
result~(\ref{contr}), we f\/ind
\begin{gather*}
e^{-i\Gamma_0}\hat\Gamma_I e^{i\Gamma_0} 
    = -\frac{\kappa}{3!}\,\Phi_{ABC}\!\!\int \!\ud^4x\! \left(
      \frac{1}{G^3}\,\bar\Pi_A\bar\Pi_B\bar\Pi_C  \!
      - \! \frac{3}{G^2}\,\bar\Pi_A\bar\Pi_B\hat{X}_C 
\! +\! \frac{3}{G}\,\bar\Pi_A\hat{X}_B\hat{X}_C
     \! -\!\hat{X}_A\hat{X}_B\hat{X}_C \right)
\end{gather*}
so that
\begin{gather*}
   \delta =\sum_{n=1}^{\infty} \kappa^n\delta_n,
\end{gather*}
represents the ef\/fective action (\ref{Gamma_eff}) as a
perturbative series in powers of $\kappa$. For instance, up to and
including the second order in $\kappa$ we have
\begin{gather}
\label{Gamma_eff2}
   \Gamma_{\eff}=i\ln\frac{A}{N}+\Gamma_0
   -i\kappa\delta_1-i\kappa^2\left(\delta_2-\frac{1}{2}\delta_1^2\right)
   -\cdots
\end{gather}
where
\begin{gather}
\label{delta1}
   \delta_1 = \frac{-i}{3!G^3}\,\Phi_{ABC}\int\ud^4x\bar\Pi_A\bar\Pi_B
              \bar\Pi_C ,
\\
   \delta_2 -\frac{\delta^2_1}{2}  = -\frac{i}{8G^5}\,
   \Phi_{ABC}\Phi_{AEF}\int\ud^4x \bar\Pi_B \bar\Pi_C \bar\Pi_E \bar\Pi_F
   \nonumber \\
\phantom{\delta_2 -\frac{\delta^2_1}{2}  =}{} +
   \delta (0) \int\ud^4x \frac{\bar\Pi_A^2}{(8G^2)^2}
   +[\delta (0)]^2 \int\ud^4x \frac{3i}{32G^3} .\label{delta2}
\end{gather}

The real factor $A(\bar\Pi )$ is always chosen such as to cancel
the imaginary part of the ef\/fective action. To the approximation
considered we have, for instance,
\begin{gather*}
   A(\bar\Pi )
    = N\,\exp\left\{\delta (0)
                    \left(\frac{\kappa}{8G^2}\right)^2 \int\ud^4x
                    \bar\Pi_A^2\right\}
 = N\left(1+ \delta(0) \frac{\kappa^2}{64G^4}
                    \int\ud^4x \bar\Pi_A^2 +\cdots \right).
\end{gather*}
It contributes to the measure of the functional integral over
$\sigma_a$, $\phi_a$.

\subsection{Giving physical meaning to the singularities}

The terms with the $\delta (0)$ function require further
explanation. The fact that auxiliary f\/ields can lead to special
problems with inf\/inities is well-known
\cite{Weinberg:1997,Bjorken:1965,Weinberg:2000,Abers:1973}. To see
in our case the origin of the encountered singularities and endow
them with physical meaning we shall use the language of Feynman
diagrams. The graphs contributing to lowest order in $\kappa$ are
shown in Fig.~1.

\begin{figure}[h]
\centerline{\includegraphics[width=5cm]{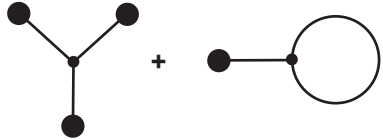}} \caption{The
lowest order graphs contributing to $\delta_1$.} \label{fig1}

\end{figure}

In these diagrams, a line segment (straight or curved) stands for
a ``propagator''
\begin{gather}
\label{propag}
   \Delta_{AB}(x-y)=-(i/G)\delta_{AB}\delta (x-y),
\end{gather}
extracted from the last exponent in equation~(\ref{Zpt2}). A
f\/illed circle at one end of a line segment corresponds to the
external f\/ield, $i\int\ud^4x\bar\Pi_A(x)$, and a vertex joining
three line segments is used for $i\kappa\Phi_{ABC}\int\ud^4x$.

The f\/irst diagram represents the $\delta_1$ term in
equation~(\ref{delta1}). The contribution of the second tadpole
diagram is equal to zero. Indeed, the vertex contains the group
factor $\Phi_{ABC}$. The contraction of any two indices in this
factor by $\delta_{AB}$ from the propagator (situation occurring
for the tadpole graph) reduces it to zero, according to
equation~(\ref{contr}). We thus f\/ind that tadpole diagrams do
not contribute due to the f\/lavour structure of the 't Hooft
interaction.

To next to leading order in $\kappa$ we have the four graphs shown
in Fig.~2.

\begin{figure}[th]
\centerline{\includegraphics[width=10cm]{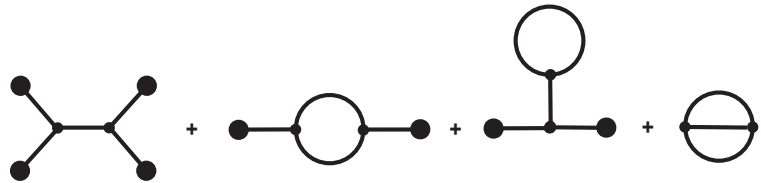}}
\caption{The diagrams of the $\kappa^2$ order contributing
         to $\delta_2$.}
\label{fig2} \vspace{-3mm}
\end{figure}

As we just learned, the third diagram does not contribute. The
other ones correspond exactly to the three terms of
equation~(\ref{delta2}). The f\/irst tree diagram is f\/inite. The
second one-loop diagram has a divergent factor $\delta (0)$. The
last two-loop diagram contributes as $\delta (0)^2$. These
singularities were caused by the local structure of the last
exponent in equation~(\ref{Zpt2}) or, which is the same, by the
$\delta (x-y)$ term in the propagator (\ref{propag}). We believe
that if one would start from non-local NJL interactions, the
singularities could be weaker or would even disappear.

The factor $\delta (0)$ requires a regularization. This is an
expected trouble in the NJL model which is nonrenormalizable and,
as a consequence, the fundamental interactions must be cut of\/f.
The cutof\/f is an ef\/fective, if crude, implementation of the
known short distance behaviour of QCD within the model. The
problem with $\delta (0)$ singularities can also be analysed using
the spectral representation method to evaluate the integral
(\ref{intJ}) \cite{Osipov:2005a}. For a one-dimensional f\/ield
theory it takes the form
\begin{gather*}
     \delta (0)_{\rm reg}=\frac{\Lambda}{2\pi}
     +\frac{1}{L} ,
\end{gather*}
where $L$ refers to the box size in which the system is put.
Therefore the second term does not contribute in the limit
$L\rightarrow\infty$. The cutof\/f $\Lambda$ cuts the density of
Fourier harmonics related to the auxiliary bosonic variables, its
f\/inite value must be f\/ixed  by confronting the model with
experiment; it may dif\/fer from the value of $\Lambda_q$ in
equation~(\ref{effpot}), which is associated with the integration
over the fermionic degrees of freedom. In this interpretation the
model is ef\/fectively f\/inite, including the higher order
corrections\footnote{Actually, if one works with the NJL model,
one must choose among several known regularizations.
Unfortunately, the dimensional regularization (DR) cannot be used.
The gap equation in this case does not have solutions and as a
result there is no dynamical chiral symmetry breaking. This is why
we cannot simply take advantage of the well-known result $\delta
(0)=0$ (in DR) to avoid the problem.}.

The diagrams are a very convenient language to understand another
feature related to the multi-loop contributions: the phase factor
corresponding to the diagram can be simply calculated. Indeed, it
is easy to see that for any diagram the formula
\begin{gather*}
   E=2I-3V
\end{gather*}
is fulf\/illed. Here $E$ is the number of external f\/ields, $I$
stands for the number of internal lines, and $V$ is the number of
vertices. On the other hand, the number of loops, $L$, is given by
\begin{gather*}
   L=I-V-E+1 .
\end{gather*}
This is because the number of loops in a diagram is equal to the
number of $\delta$ functions surviving after all integrations over
coordinates are performed, except for one over-all integration
related to the ef\/fective action. Every internal line contributes
one $\delta$ function, but every vertex or external f\/ield
carries an integration over the corresponding coordinate, and thus
reduces the number of $\delta$ functions by one.

This result shows that the overall phase factor of a given graph
$i^{I-E-V}=i^{L-1}$ is entirely determined by the number of loops.
In particular, diagrams with an even number of loops contribute to
the ef\/fective action (the argument of ${\cal Z}$), while the
diagrams with odd number of loops contribute to the measure (the
modulus of ${\cal Z}$), see equation~(\ref{Zpt3}).

\section[The loop expansion of ${\cal Z}$]{The loop expansion of $\boldsymbol{\cal Z}$}

The perturbative series, considered in the previous section, can
be resummed. This is done by introducing a parameter t in the
perturbative representation (\ref{Zpt1}) of the functional
integral according to the substitution:
\begin{gather*}
G,\kappa,\Pi_A,\Delta_A \rightarrow tG,t\kappa,t\Pi_A,t\Delta_A.
\end{gather*}
Then equation~(\ref{Gamma_eff2}) to second order in $\kappa^2$
becomes
\begin{gather*}
\Gamma_{\eff}\rightarrow t \Gamma'_{\eff}(t),
\\
   \Gamma'_{\eff}(t)=\frac{i}{t}\ln\frac{A({\bar{\Pi}_A})}{N}+\Gamma_0
   -i\kappa\delta_1-i\kappa^2\Bigg[-\frac{i}{8G^5}\,
   \Phi_{ABC}\Phi_{AEF}\int\ud^4x \bar\Pi_B \bar\Pi_C \bar\Pi_E \bar\Pi_F
\nonumber \\
\phantom{\Gamma'_{\eff}(t)=}{} +
   \frac{1}{t}\delta (0) \int\ud^4x \frac{\bar\Pi_A^2}{(8G^2)^2}
   +\frac{1}{t^2}[\delta (0)]^2 \int\ud^4x \frac{3i}{32G^3}\Bigg]
   -\cdots.
\end{gather*}
The groups in terms of inverse powers of t of $\Gamma'_{\eff}(t)$
def\/ine the new series: f\/irst all diagrams with no closed loops
(tree graphs) will carry the weight~$1$. The tree graphs yield the
result of~\cite{Reinhardt:1988}. All the one-loop diagrams (and
higher odd numbers of loop diagrams) contribute to the imaginary
part of the action, as discussed in the previous section; as
mentioned before, they will be cancelled by the appropriate choice
of the real quantity $\frac{1}{t} A({\bar{\Pi}_A})$, with $A$
being of order~$1$.  They contribute therefore to the measure of
the integral over $\Pi_A$ and were obtained in~\cite{Osipov:2004},
see also equations (\ref{I0}), (\ref{Zloop4}) below. The set of
graphs with $2n$ loops have the factor $t^{-2n}$ and contribute to
the ef\/fective action; in general they include, as a subset, all
graphs of $k^{n+1}$th order or higher in this coupling constant.
The resummed series is equivalent to the well-known loop
expansion~\cite{Coleman:1973}. The latter can be written formally
as a SPA integral which includes only the regular critical point
(SPAr), and constitutes the most economic way to obtain the  loop
corrections. A detailed derivation of the equivalence of the
resummation of the perturbative series, the loop expansion and the
SPAr is given in Section~3 of~\cite{Osipov:2005a} for a
one-dimensional analogue of the considered functional.

We proceed therefore to obtain the NLO to the ef\/fective action
through the SPAr method.
 Consider the ef\/fective mesonic action generated by the functional
\begin{gather}
     {\cal Z}[\Pi ,\Delta ]\sim
     {\cal N}
     \exp\left( i\int\ud^4x {\cal L}_\st^{(i=1)}\right)
     \int\limits^{+\infty}_{-\infty}\prod_A{\cal D}\bar{R}_A\
     \exp\left(\frac{i}{2}\int\ud^4x{\cal L}_{AB}''({\cal R}^{(i=1)})
     \bar{R}_A \bar{R}_B \right) \nonumber \\
 \phantom{{\cal Z}[\Pi ,\Delta ]\sim}{}\times  \sum_{n=0}^\infty \frac{1}{n!}
     \left(i\frac{\kappa}{3!}\ \Phi_{ABC}\int\ud^4x
     \bar{R}_A\bar{R}_B\bar{R}_C\right)^n   \label{Zloop1}
\end{gather}
which in comparison with equation~(\ref{intJisp}) has only one
critical point, related to the stable conf\/iguration, which is
the solution with $i=1$ in equations~(\ref{hsu3}). We shall
identify ${\cal L}_\st^{(i=1)}={\cal L}_\st$, and ${\cal
L}_{AB}''({\cal R}^{(i=1)})={\cal L}_{AB}''$ in the following.

By replacing the continuum of spacetime positions with a discrete
lattice of points surrounded by separate regions of small
spacetime volume $\Omega$, the functional integral (\ref{Zloop1})
may be reexpressed as a Gaussian multiple integral over a f\/inite
number of real variables $R_A(x)$ for a f\/ixed spacetime point
$x$. We think of ${\cal D}R_A$ as the inf\/inite product ${\cal
D}R_A\to\prod_x \ud R_A(x)$, $\int\ud^4x\to\Omega\sum_x$.

After evaluating the Gaussian integrals one obtains
\begin{gather}
\label{Zloop3}
     {\cal Z}[\Pi ,\Delta ] \sim
     {\cal N} \prod_x \left\{\,
     I_0\, \exp\left( i \Omega {\cal L}_\st\right)
     \left( 1 +i\frac{\kappa^2\Phi_{ABC}\Phi_{DEF}
     \delta_{ABCDEF}}{72\Omega N(N+2)(N+4)}{\cal M}
     +\cdots\right)\right\}.
\end{gather}
Here
\begin{gather*}
   {\cal M} =
   \big(\mbox{tr}\,{\cal L}''{}^{-1}\big)^3
   + 6\,\mbox{tr}\, {\cal L}''{}^{-1}
   \mbox{tr}\,\big({\cal L}''{}^{-1}\big)^2
   + 8\,\mbox{tr}\,\big({\cal L}''{}^{-1}\big)^3,
\end{gather*}
contains the complete information on the two-loop term and the
dots mean the terms corresponding to the three-loop contribution
and higher. The $I_0$ is the one-loop contribution
\begin{gather}
\label{I0}
   I_0=\frac{1}{\sqrt{\det {\cal L}''}}
       \left(\frac{2\pi}{\Omega}\right)^\frac{N}{2}
       \exp\left(i\frac{\pi}{4}\sum\limits_{j=1}^N
       \mbox{sgn}(\lambda_j)\right)
\end{gather}
where $\lambda_j$ are eigenvalues of the $N\times N$ matrix ${\cal
  L}''_{AB}$. In our case $N=18$, related to the 18-component object $R_A$, equation~(\ref{sdLr}), and
 ${\cal L}''{}^{-1}$ denotes
the inverse matrix of ${\cal L}''$. The totally symmetric symbol
$\delta_{ABCDEF}$ generalizes an ordinary Kronecker delta symbol
$\delta_{AB}$ by the recurrent relation
\begin{gather*}
   \delta_{ABCDEF}= \delta_{AB}\delta_{CDEF}+
                    \delta_{AC}\delta_{BDEF}+\cdots
                   +\delta_{AF}\delta_{BCDE} .
\end{gather*}

Up to the given accuracy we have in equation~(\ref{Zloop3})
\begin{gather}
     \prod_x \left(\, I_0\, e^{i \Omega {\cal L}_\st}
     \left( 1 +iF \right)\right) =
     \prod_x \left(\, I_0\, e^{i (\Omega {\cal L}_\st +F)}\right)
     \nonumber\\
\qquad{}     = \left(\prod_x I_0 \right) \left( \prod_x
     e^{i (\Omega {\cal L}_\st +F)}\right)
     = \left(\prod_x I_0 \right)
     e^{i\Omega\sum_x ({\cal L}_\st +F/\Omega )}\label{Zloop4}
\end{gather}
with the quantity $I_0$ containing the one-loop corrections to the
measure. It shows that in the continuum limit the two-loop
correction contributes to the ef\/fective Lagrangian as
\begin{gather}
\label{2loop}
   {\cal L}_\eff = {\cal L}_\st +
   \frac{3\kappa^2 [\delta (0)]^2 {\cal M}}{32N(N+2)(N+4)}\, .
\end{gather}
This is our f\/inal expression for the ef\/fective Lagrangian in
the two-loop approximation.

The f\/ield-dependent factor ${\cal M}$ contains all possible
mesonic vertices, including the $\sigma$-tadpole contribution to
the gap equation, contributions to the masses of scalar and
pseudoscalar nonets, as well as interaction terms. Its dependence
on the parameters $\kappa$, $G$, $m$, $\hat {m}$ enter through the
expansion coef\/f\/icients $H_A^{(i)}$, equations (\ref{HA}),
(\ref{saddle-1}).

Let us do some estimates to justify the result. For this purpose
let us simplify the integ\-ral~(\ref{Zloop1}). After neglecting
the symmetry group and discretizing the spacetime it takes the
form
\begin{gather}
\label{Zls}
     {\cal Z}[\Pi ,\Delta ] \sim \prod_x \int\ud R_x
     \exp\left\{ i \Omega \left( {\cal L}_{\st} +\frac{1}{2}
     {\cal L}_\st''R_x^2 + \frac{1}{3!}{\cal L}_\st'''R_x^3
     \right)\right\} .
\end{gather}
To justify the stationary phase approximation for the integral
(\ref{Zls}) we assume that
\begin{gather}
\label{sclc}
     \Omega {\cal L}_\st \gg 1.
\end{gather}
The dominating role of the Gaussian integral is ref\/lected in the
fact that essential values for $R_x$ in the integral have the
order $R_x^2\sim 1/(\Omega {\cal L}_\st'')$. For the cubic term it
follows then
\begin{gather*}
     \Omega {\cal L}_\st'''R_x^3\sim \sqrt{
     \frac{({\cal L}_\st''')^2}{\Omega ({\cal L}_\st'')^3}}
     \sim\sqrt{\frac{\kappa^2}{\Omega G^3}} \sim\sqrt{\zeta} ,
\end{gather*}
where we have used that in the model considered here, ${\cal
L}_\st'''\sim\kappa$, ${\cal L}_\st''\sim G$. If the parameters of
the model can be chosen in such a way that the inequality
$\zeta\ll 1$ is fulf\/illed, the cubic power of $R_x$ yields terms
that go to zero relative to the Gaussian term as $\zeta\to 0$, and
the stationary phase approximation will be justif\/ied. Note that
$\Omega$ may be written as an ultraviolet divergent integral
regularized by introducing a cutof\/f $\Lambda$
\begin{gather*}
   \Omega^{-1} = \delta^4(0)\sim
                 \int_{-\Lambda /2}^{\Lambda /2}
                 \frac{\ud^4k_\e}{(2\pi )^4}=
                 \left(\frac{\Lambda}{2\pi}\right)^4.
\end{gather*}
Therefore, the inequality restricts the value of $\Lambda$ from
above.

Meanwhile, it is interesting to see that the inequality $\zeta\ll
1$ is an exact equivalent of (\ref{sclc}). Indeed, in the
essential region, i.e., around a sharp minimum, one has ${\cal
L}_\st'\sim R_x {\cal L}_\st$, ${\cal L}_\st''\sim R_x^2{\cal
L}_\st$, and so on, thus
\begin{gather*}
     \zeta\sim
     \frac{({\cal L}_\st''')^2}{\Omega ({\cal L}_\st'')^3}
     \sim\frac{({R_x^3\cal L}_\st )^2}{\Omega (R_x^2{\cal L}_\st )^3}
     \sim\frac{1}{\Omega {\cal L}_\st} .
\end{gather*}

Therefore, the asymptotical series (\ref{Zloop3}) with the
ultraviolet cutof\/f imposed in the continuum limit is sensible.
One deals here, actually, with a series in powers of the
dimensionless parameter~$\zeta$. The expansion is formally
justif\/ied for $\zeta\ll 1$.

\subsection[The NLO effective potential]{The NLO ef\/fective potential}

As an example let us obtain the NLO corrections due to
(\ref{2loop}) to the ef\/fective potential for the $SU(3)$ limit
and with the current quark masses set to zero. Here only the
regular critical point was taken into account. For the case of the
full SPA, one has to consider the same expression also for the
singular point. The term linear in the scalar f\/ields coming from
the two-loop correction is calculated with the help of
``Mathematica'' \cite{Math}
\begin{gather*}
\frac{1}{2}\sum _i c_{i,a} \sigma_a =
(\sigma_u+\sigma_d+\sigma_s)\sum_i
\frac{-6 \kappa}{G^5} \omega_i \gamma \nonumber \\
\phantom{\frac{1}{2}\sum _i c_{i,a} \sigma_a =}{}\times
\frac{\big(121 - 1552 \omega_i^2 + 7832 \omega_i^4 - 17980
\omega_i^6 + 14467 \omega_i^8 +
       3592 \omega_i^{10}\big)}{(1 - 4 \omega_i^2)^4 (1 - \omega_i^2)^4(1+2\omega_i)}
\end{gather*}
and it must be added to the leading order term
equation~(\ref{gec1}). Here the coef\/f\/icient
\begin{gather*}
\gamma= \left[\frac{\Lambda}{2 \pi}\right]^8 \frac{3 \kappa^2}{32
N (N+2)(N+4)},
\end{gather*}
and $\omega_i=\omega_{u,i}=\omega_{d,i}=\omega_{s,i}=\frac{k
h_i}{16 G}$ with the index $i$ denoting the two possible solutions
for $h_i$
 to the SPA equations in the $SU(3)$ case, equation~(\ref{hsu3}).
One must integrate $c_i$ with respect to the quark mass to obtain
the correction to the ef\/fective potential, since
\begin{gather*}
dV_{\rm corr}=\frac{1}{2}\sum_i \frac{\partial V^i_{\rm
corr}}{\partial m(i)} dm(i) = \frac{1}{2}\sum_i c_i dm(i)=-\sum_i
c_i \frac{8 G^2}{\kappa} (1+2\omega_i) d\omega_i,
\end{gather*}
where for the last equality the equations~(\ref{saddle-1}) were
used to obtain
\begin{gather*}
dm(i)=-\left(G+\frac{\kappa h_{i}}{8} \right) dh_{i}=-\frac{16
G^2}{\kappa} (1+2\omega_{i}) d\omega_{i}.
\end{gather*}
Here $m=m_u=m_d=m_s$  and $m(i)=m$ for both solutions. With
$x_i=\omega_i^2$ one gets
\begin{gather*}
dV_{\rm corr}=\frac{48}{G^3} \gamma  \sum_i \left[\frac{\big(121 -
1552 x_i + 7832 x_i^2 - 17980 x_i^3 + 14467 x_i^4 +
        3592 x_i^5\big)}{(1 - 4 x_i)^4 (1 - x_i)^4}\right]dx_i.
\end{gather*}
As a result we obtain
\begin{gather*}
V_{\rm corr}=\frac{48}{G^3} \gamma  \sum_i\frac{55-583 x_i+2023
x_i^2-2037 x_i^3-898 x_i^4}{2(1-x_i)^3 (1-4 x_i)^3}.
\end{gather*}
Therefore the complete expression for the $SU(3)$ ef\/fective
potential is
\begin{gather*}
U(m)=\frac{12 G m}{\kappa} -\frac{3 N_c}{16 \pi^2} \left[m^2
\left(\Lambda_q^2-m^2 \ln
\left(1+\frac{\Lambda_q^2}{m^2}\right)\right)+\Lambda_q^4 \ln
\left(1+\frac{m^2}{\Lambda_q^2}\right)\right] +V_{\rm corr}(m).
\end{gather*}
There will be poles at $x_i=\{\frac{1}{4},1\}$. In terms of $m$
there can be in principle up to 4 non-degenerate poles altogether,
 two for each critical point. They are due to the zero eigenvalues of ${\cal L}''$,
  which appear for certain values of the coupling constants \cite{Osipov:2004}
  in subsets of the 18-f\/ield conf\/igurations. In the neighborhood of these
  points the SPA method is unreliable. This puts constraints on the coupling
  constants, which must be chosen such that the poles appear far from the region
  of physical interest (the quark mass value at the local minimum of the ef\/fective potential).
  One sees however that the solution $h_2$, related with the singular critical point,
  has always a pole at $m=0$, since for this case $x_2=1$. It is one more indication
  that the contribution of the singular point must be excluded from the analysis.
  Only the perturbative regime is sensible.
In this case the ef\/fective potential reads
\begin{gather*}
U_1(m)=\frac{1}{4}
\left(3 G h_1^2+\frac{\kappa}{4} h_1^2\right)\\
\phantom{U_1(m)=}{}-\frac{3 N_c}{16 \pi^2} \left[m^2
\left(\Lambda_q^2-m^2 \ln \left(1+\frac{\Lambda_q^2}{m^2}\right)
\right)+\Lambda_q^4
\ln\left(1+\frac{m^2}{\Lambda_q^2}\right)\right] +V_{1,{\rm
corr}}(m)
\end{gather*}
where the index $1$ labels the contribution of the regular point.
The solution $h_1$ will lead to poles at $m=\frac{12 G^2}{\kappa}$
and $m=\frac{32 G^2}{\kappa}$. The parameters $G$, $\kappa$ can be
chosen such that the pole positions are far away from the
constituent quark mass solution of the gap equation. For instance
for $G=13.5$ GeV$^{-2}$, $\kappa=-1300$ GeV$^{-5}$, $\Lambda=1.8$
GeV, $\Lambda_q=0.82$ GeV the two poles due to $h_1$ appear at
$m\sim 1.7$ GeV and  $m\sim 4.5$ GeV. For $G=2.15$ GeV$^{-2}$,
$\kappa=-53$ GeV$^{-5}$, $\Lambda=1.9$ GeV, $\Lambda_q=1.64$ GeV
one has the two poles at $m\sim 1.$ GeV amd $m\sim 2.8$ GeV. Both
sets of parameters lead to good meson mass spectra and weak decay
constants in the realistic case with $SU(2)\times U_A(1)$
symmetry~\cite{Osipov:2006}. In conclusion, from the point of view
of the complete SPA (inclusion of regular and singular points),
there is no improvement, because of the bad pole at zero. From the
point of view of the perturbative (regular) solution, the poles
are far away on the positive m axis and the present approach can
be used to study systematically NLO corrections.

\subsection*{Acknowledgements}

B. Hiller is very grateful to the Organizers of the sixth
International Conference on ``Symmetry in Nonlinear Mathematical
Physics'', Kyiv, June 20--26, 2005, for the kind invitation and
for the excellent organization which gave rise to a very
interesting and pleasant meeting. This work has been supported by
grants provided by Funda\c c\~ao para a Ci\^encia e a Tecnologia,
POCTI/FNU/50336/2003 and POCI/FP/63412/2005; This research is part
of the EU integrated infrastructure initiative Hadron Physics
project under contract No.RII3-CT-2004-506078. A.A.~Osipov also
gratefully acknowledges the Funda{\c{c}\~{a}}o
 Calouste Gulbenkian for f\/inancial support.
B.~Hiller and A.A.~Osipov are very grateful to V.~Miransky and
V.N.~Pervushin for discussions.

\LastPageEnding

\end{document}